\documentclass[useAMS,usenatbib]{mn2e}
\usepackage{times}
\input{epsf}

\title[]
{An additional soft X-ray component in the dim low/hard state of
black hole binaries}

\author[C.Y. Chiang, Chris Done, M. Still, and O. Godet]
{C. Y. Chiang$^{1}\thanks{E-mail: chia-ying.chiang@durham.ac.uk}$, Chris Done$^{1}$, M. Still$^{2,4}$, and O. Godet$^{3}$\\
$^1$Department of Physics, University of Durham, South Road, Durham
DH1 3LE, UK\\
$^2$Mullard Space Science Laboratory, Dorking, Surrey, UK\\
$^3$X-ray and Observational Astronomy Group, Department of Physics and Astronomy, University of Leicester, LE1 7RH, UK\\
$^4$NASA Ames Research Center, Moffett Field, CA 93045, USA\\}

\date{Submitted to MNRAS}
\pagerange{\pageref{firstpage}--\pageref{lastpage}} \pubyear{2009}

\begin{document}

\topmargin = -0.5cm

\maketitle

\label{firstpage}

\begin{abstract}\

We test the truncated disc models using multiwavelength
(optical/UV/X-ray) data from the 2005 hard state outburst of the black
hole SWIFT J1753.5-0127.  This system is both fairly bright and has
fairly low interstellar absorption, so gives one of the best datasets
to study the weak, cool disc emission in this state.  We fit these
data using models of an X-ray illuminated disc to constrain the inner
disc radius throughout the outburst.  Close to the peak, the observed
soft X-ray component is consistent with being produced by the inner
disc, with its intrinsic emission enhanced in temperature and
luminosity by reprocessing of hard X-ray illumination in an overlap
region between the disc and corona. This disc emission provides the
seed photons for Compton scattering to produce the hard X-ray
spectrum, and these hard X-rays also illuminate the outer disc,
producing the optical emission by reprocessing.

However, the situation is very different as the outburst declines.
The optical is probably cyclo-synchrotron radiation, self-generated by
the flow, rather than tracing the outer disc. Similarly, limits from
reprocessing make it unlikely that the soft X-rays are directly tracing the
inner disc radius. Instead they appear to be from a new component. This is
seen more clearly in a similarly dim low/hard state spectrum from XTE
J1118+480, where the 10$\times$ lower interstellar absorption allows a
correspondingly better view of the UV/EUV emission. The very small
emitting area implied by the relatively high temperature soft X-ray
component is completely inconsistent with the much larger, cooler, UV
component which is well fit by a truncated disc. We speculate on the
origin of this component, but its existence as a clearly separate
spectral component from the truncated disc in XTE J1118+480 shows that it does
not simply trace the inner disc radius, so cannot constrain the
truncated disc models.

\end{abstract}


%==============================================
\section{Introduction} \label{sec:introduction}\
%==============================================

The accretion disc in black hole binary systems (BHB) is unstable at
the point where Hydrogen goes from being predominantly neutral to
ionised. This gives rise to dramatic, transient outbursts in which
the reservoir of material built up in the outer disc during
quiescence can accrete down onto the black hole. The peak luminosity
of the outburst is set by the mass in the quiescent disc and the
time taken for this to accrete. Both these depend on the size of the
disc taking part in the outburst, which is set by the binary
separation. Hence there is a link between orbital period and peak
luminosity, with short period systems showing smaller peak outburst
luminosities (\citealt{KR98}; \citealt{Shahbaz 98}; \citealt{Lasota 01};
\citealt{DGK07} hereafter DGK07).

The peak luminosity determines the number of spectral states which are
seen during the outburst. Systems with peak luminosity below $\sim 0.1
L_{Edd}$ remain in the low/hard state (LHS), where the energy output
peaks at $\sim$ 100 keV. Brighter systems instead show a distinct
transition to a much softer thermal dominated state (TDS, also
termed a high/soft state) which peaks at $\sim$ 1 keV, while
the brightest systems can also show a very high state (VHS,
alternatively steep power law state). These dramatic changes in the
spectrum are correlated with equally dramatic changes in the rapid
variability properties (see e.g. \citealt{MR06}) and the jet
\citep{FBG04}, implying distinct changes in the nature and geometry of
the accretion flow. These can be plausibly explained if there is a
hot, optically thin, geometrically thick solution to the accretion
flow equations at low luminosities (\citealt{SLE76};
\citealt{NY95}).This can be put together with a cool, optically thick,
geometrically thin disc \citep{SS73} in the truncated disc/hot flow
model where the disc progressively replaces more of the inner hot flow
as the mass accretion rate increases.  The distinct hard/soft spectral
transition then marks the point at which the cool disc extends down to
the last stable orbit, replacing all the hot flow (\citealt{EMN97};
DGK07).

Such models make a clear prediction that the inner radius of the cool
disc should recede as the LHS drops in luminosity. This can in
principle be tested using the colour temperature and luminosity of the
cool disc component to estimate its radius
(e.g. \citealt{Poutanen97}; \citealt{ST99}). However, such observations
are complicated by the fact that the disc is at low temperature at low
luminosities, $\sim 0.3$~keV for a $10M_\odot$ black hole at $0.02
L_{Edd}$, even if the disc extends down to $6R_g$ (where
$R_g=GM/c^2$).  Such low energy X-rays cannot be seen with the 3~keV
bandpass limit of \emph{RXTE}, the satellite which has accumulated the
most BHB data to date (\citealt{DG03}; \citealt{Dunn09}).  CCD
observations can extend the bandpass down to lower energies, but these
are still not straightforward to interpret as the disc is not the
dominant spectral component in this state. Instead, the hard X-ray
emission dominates the spectrum, so irradiation can change the disc
temperature (\citealt{GDP08}, hereafter GDP08), while
the disc luminosity can be underestimated due to photons which are
Compton upscattered into the hard X-ray spectrum \citep{Makishima08}.
This component is still not easy to study even with CCD
detectors since low energy X-rays are absorbed by interstellar
gas. Most BHB are in the galactic plane so have gas columns of $N_H\ge
10^{22}$~cm$^{-2}$ which effectively block emission below 1~keV.

However, there are a few black holes which have intrinsically much
lower columns which have been observed with CCD detectors, namely XTE
J1118+480: $N_H\sim 1.1\times 10^{20}$~cm$^{-2}$, XTE J1817-330:
$1.1\times 10^{21}$~cm$^{-2}$ and SWIFT1753.5-0127: $2-3\times
10^{21}$~cm$^{-2}$ \citep{Cabanac09}. XTE J1118+480 has only
snapshot spectra available (\citealt{Hynes 2000}; \citealt{Esin 01};
\citealt{Frontera 01}; 2003), but both SWIFT1753.5-0127 and XTE
J1817-330 were well sampled throughout their outbursts by the \emph{Swift}
satellite. However, XTE J1817-330 is a long period system, so spends
most of its time in the disc dominated state, with only two, rather
faint, LHS spectra at the end of its outburst. Thus the short period
LHS outburst of SWIFT1753.5-0127 is the best candidate to study the
disc evolution in the LHS.  Here we analyse these data with
sophisticated models to describe the behaviour of the X-ray irradiated
disc (GDP08; \citealt{GDP09}, hereafter GDP09) to constrain its inner radius as the outburst declines.

%========================================
\section{Data reduction} \label{sec:data}
%========================================
\subsection{\emph{Swift}}\
\begin{figure}
\begin{center}
\leavevmode \epsfxsize=8.5cm \epsfbox{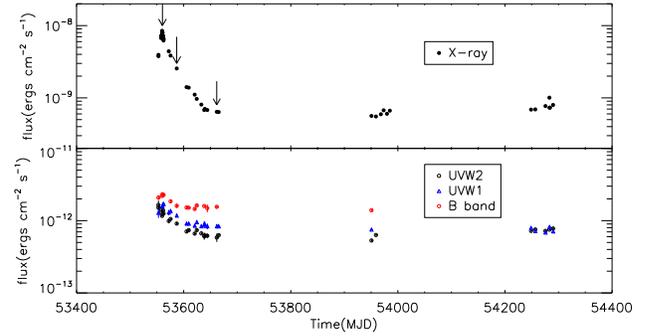}
\end{center}
\caption{The top panel shows the evolution of the observed flux in the
  \emph{Swift} XRT band, while the lower panel shows the simultaneous
UVOT data (in UVW1, UVW2 and B filters, where available).}
\label{flux}
\end{figure}

\begin{figure}
\begin{center}
\leavevmode \epsfxsize=8.5cm \epsfbox{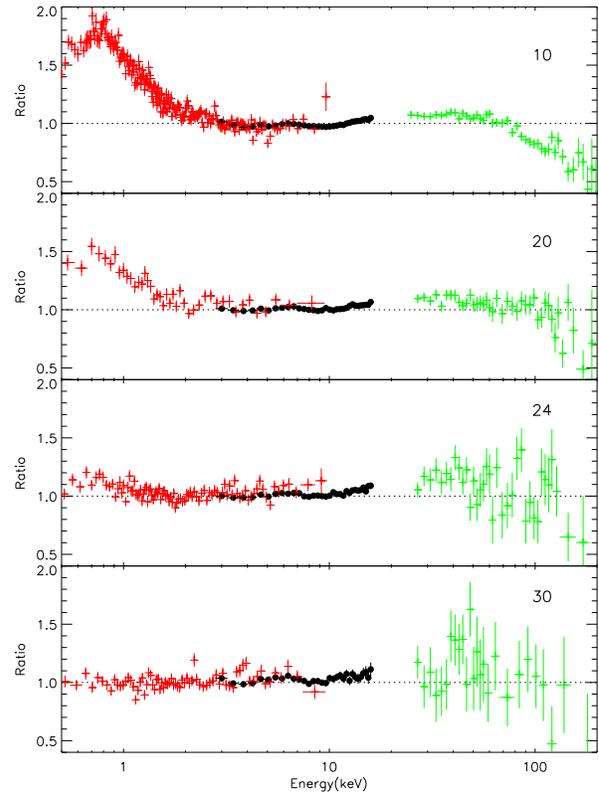}
\end{center}
\caption{The ratio of the X-ray data (red:XRT, black:PCA, green:HEXTE)
to an absorbed power law model fit to the 2-4 and 7-10~keV 'continuum
dominated' bands for spectra taken from the peak (10) through to the
end of the decline (30) as marked by arrows on Fig 1. At the peak
is clear that the continuum spectrum strongly declines above 70~keV,
indicating that the electron temperature of the thermal Comptonisation
is low enough to be seen. There is also a strong additional soft
component at low energies, most probably from the accretion disc,
together with a small features around the iron line and edge,
indicating reflection of the continuum from the accretion disc.
As the outburst declines, the electron temperature increases, so the
high energy break can no longer be seen, and the soft X-ray emission drops
dramatically in strength. }
\label{residual}
\end{figure}

\begin{table*}
\begin{minipage}{160mm}
 \caption{The table details the datasets we use from \emph{Swift} and
\emph{RXTE}.  We refer to the combined spectrum by the last two
numbers of its \emph{Swift} obsID. The remaining columns give the QPO
frequency from fitting the \emph{RXTE} power spectrum, followed by the
results for fitting the 0.5-200~keV X-ray data with a {\tt
wabs*(diskbb+thCompml)} model. We fix the absorption to $0.2\times
10^{22}$~cm$^{-2}$, and assume reflection is from neutral, solar
abundance material inclined at $60^\circ$. We tie the seed photons for
Comptonisation to the disc temperature in these fits, but then also
show the difference in $chi^2$ for removing the disc and fixing the
seed photon energy for the Compton scattering to 0.1~eV }
\label{bigtable}
\begin{tabular}{@{}ccccccccccr}
\hline\hline

 Number & \emph{Swift} & \emph{RXTE}  & QPO (Hz) & $kT_{disc}(keV)$ & $N_{dbb}$ &$\Gamma$ & $\Omega/2\pi$ & $N_{pl}$ & $\chi^{2}/\nu$ & $\bigtriangleup\chi^{2}$ \\
\hline
 00 & 00143778000   &91094-01-01-00 & $0.64\pm0.01$  & $0.22\pm0.01$   & 47557      & $1.80\pm0.02$  & $0.34^{+0.10}_{-0.13}$
    & 0.58   & 712/713   & 699\\
 03 & 00030090003   &91094-01-01-04 & $0.83\pm0.01$ & $0.25^{+0.00}_{-0.01}$   & 63806      & $1.86^{+0.00}_{-0.02}$ & $0.35^{+0.05}_{-0.08}$
    & 1.06   & 544/548   & 1024\\
 04 & 00030090004   &91423-01-01-04 & $0.91\pm0.01$  & $0.24^{+0.00}_{-0.01}$   & 66358      & $1.86\pm0.01$ & $0.27\pm0.06$
    & 1.06   & 870/757   & 1503\\
 07 & 00030090007   &91094-01-02-01  & $0.75^{+0.01}_{-0.02}$  & $0.24^{+0.01}_{-0.00}$   & 69967      & $1.83^{+0.02}_{-0.01}$ & $0.36^{+0.04}_{-0.08}$
    & 1.16   & 702/679   & 1112\\
 06 & 00030090006   &91094-01-02-01  & ``  & $0.24^{+0.00}_{-0.01}$   & 83839      & $1.82^{+0.02}_{-0.01}$ & $0.32^{+0.08}_{-0.07}$
    & 1.26   & 868/762   & 1674\\
 10 & 00030090010   &91094-01-02-00  & $0.72^{+0.00}_{-0.01}$ & $0.25^{+0.00}_{-0.01}$   & 70377      & $1.82^{+0.01}_{-0.02}$ & $0.33^{+0.03}_{-0.04}$
    & 1.17   & 752/726   & 2192\\
 08 & 00030090008   &91094-01-02-00 & `` & $0.28\pm0.02$   & 36514      & $1.81\pm0.01$ & $0.28^{+0.05}_{-0.04}$
    & 0.86   & 275/273   & 232\\
 09 & 00030090009   &91094-01-02-00  & `` & $0.24^{+0.01}_{-0.00}$   & 73119      & $1.81^{+0.01}_{-0.01}$ & $0.28\pm0.05$
    & 1.17   & 873/784   & 2955\\
 11 & 00030090011   &91094-01-02-02   & $0.70\pm0.01$ & $0.25\pm0.02$   & 59219      & $1.81\pm0.01$ & $0.29\pm0.06$
    & 1.02   & 333/320   & 250\\
 12 & 00030090012   &91094-01-02-02 & ``  & $0.23^{+0.01}_{-0.00}$   & 79673      & $1.81\pm0.01$ & $0.29^{+0.05}_{-0.06}$
    & 1.09   & 860/746   & 1898\\
 13 & 00030090013   &91094-01-02-02  & `` & $0.23^{+0.01}_{-0.00}$   & 75104      & $1.81\pm0.01$ & $0.30\pm0.06$
    & 1.08   & 622/633   & 1134\\
 15 & 00030090015   &91094-01-02-03  & $0.63\pm0.01$  & $0.23^{+0.00}_{-0.01}$   & 69620      & $1.79\pm0.01$ & $0.31^{+0.07}_{-0.06}$
    & 0.99   & 818/743   & 1231\\
 16 & 00030090016   &91094-01-02-03  & `` & $0.25\pm0.01$   & 49618      & $1.79\pm0.01$ & $0.31^{+0.06}_{-0.07}$
    & 0.94   & 717/655   & 1030\\
 18 & 00030090018   &91423-01-03-06 & $0.47\pm0.01$  & $0.23^{+0.01}_{-0.02}$   & 42447      & $1.73\pm0.01$ & $0.24\pm0.06$
    & 0.64   & 396/435   & 347\\
 19 & 00030090019   &91423-01-04-02 & $0.45^{+0.00}_{-0.01}$  & $0.22\pm0.01$   & 39291      & $1.72\pm0.01$ & $0.28^{+0.07}_{-0.06}$
    & 0.57   & 626/646   & 732\\
 20 & 00030090020   &91423-01-06-00 & $0.29\pm0.01$  & $0.23^{+0.01}_{-0.02}$   & 17415      & $1.67^{+0.02}_{-0.01}$ & $0.20^{+0.07}_{-0.09}$
    & 0.36   & 433/460   & 178\\
 21 & 00030090021   &91423-01-08-02 & -  & $0.22\pm0.02$   & 9122       & $1.65\pm0.03$ & $0.19^{+0.16}_{-0.17}$
    & 0.21   & 514/518   & 50\\
 23 & 00030090023   &91423-01-10-00  & - & $0.17\pm0.02$   & 14876      & $1.65^{+0.01}_{-0.02}$ & $0.16^{+0.05}_{-0.09}$
    & 0.18   & 635/659   & 32\\
 24 & 00030090024   &91423-01-11-00  & - & $0.16^{+0.02}_{-0.03}$   & 16135      & $1.67^{+0.01}_{-0.02}$ & $0.24^{+0.11}_{-0.13}$
    & 0.16   & 602/631   & 19\\
 26 & 00030090026   &91423-01-13-00 & -  & $0.17^{+0.05}_{-0.10}$   & 4629       & $1.63^{+0.02}_{-0.04}$ & $0.13^{+0.14}_{-0.13}$
    & 0.11   & 559/562   & 4\\
 30 & 00030090030   &91423-01-16-01 & -  & $0.04^{+0.05}_{-0.03}$   & 934        & $1.64^{+0.01}_{-0.02}$ & $0.09^{+0.14}_{-0.09}$
    & 0.11   & 559/592   & 3\\
 31 & 00030090031   &91423-01-17-00 & -   & $0.13^{+0.04}_{-0.12}$   & 13157      & $1.63^{+0.02}_{-0.03}$ & $0.14^{+0.16}_{-0.14}$
    & 0.10   & 576/508   & 3\\
\hline\hline\\
\end{tabular}
\end{minipage}
\end{table*}

We use publicly available \emph{Swift} data from SWIFT J1753.5-0127
taken during the period from July 2005 to July 2007. We extracted both
X-ray telescope (XRT) and UVOT data. All of our XRT data was in
Windowed Timing mode, and we extracted source counts using a circle
with radius of 20 pixels.  Background was taken from an off source
region using a circle of the same size.  However, the source was very
bright during the peak of the outburst, and some of the datasets were
piled up as the count rate was above 100 counts $s^{-1}$.  We
determined the size of the central piled up region by excluding
progressively larger radii regions until the spectra stopped softening
at high energies.  This gave an exclusion region of 3 pixels for
brightest observations, and 2 pixels for more moderately piled up
data.  The data was grouped to 20 counts $s^{-1}$, and fit between
0.5-10 keV.

The evolution of the observed (i.e. not corrected for interstellar
absorption) flux in the X-ray (upper) and UV (lower) bands is plotted
in Fig. \ref{flux}. We derive the flux by integrating the absorbed model fit
over 0.5-10~keV for the X-rays.  For UV data, we find energy band for
each filter and define the effective bandpass to be between the
energies determined by FWHM ($5.6\times10^{-3}$ to $7.45\times10^{-3}$
keV for UVW2 filter, $4.2\times10^{-3}$ to $5.45\times10^{-3}$ keV for
UVW1 filter, and $2.55\times10^{-3}$ to $3.2\times10^{-3}$ keV for B
filter).

The observations fall into 3 clearly distinct time segments.  The
first covers the outburst rise and decline, after which the flux
remains fairly constant.  Hence in this paper we concentrate on this
first data group, in order to track the evolution of the disc during
the flux decline.

\subsection{\emph{RXTE}}\

The \emph{RXTE} satellite also followed the 2005 outburst. We use the
standard extraction techniques with the bright source background to
derive PCA spectra from all layers of detector 2, adding 1 percent
systematic error and use these data from 3-16 keV.  We also extract
Standard 1 power spectra over the full 2-60~keV bandpass for intervals
of 256~s, and fit these with multiple Lorentzian components in order
to determine the QPO frequency. These are given in Table
\ref{bigtable}.

We extract HEXTE data from cluster 0 and use these over the energy
range 25-200 keV. We select the subset of \emph{Swift} and
\emph{RXTE} data which are taken within one day of each other.  This
gives a total of 27 observations which have quasi-simultaneous
coverage of the optical/UV/X-ray/hard X-ray spectrum.
Table \ref{bigtable} details these composite spectra.

%====================================
\section{Disc plus Comptonisation plus reflection} \label{sec:results}
%====================================

\subsection{X-ray data}
\begin{figure}
\begin{center}
\leavevmode \epsfxsize=8.5cm \epsfbox{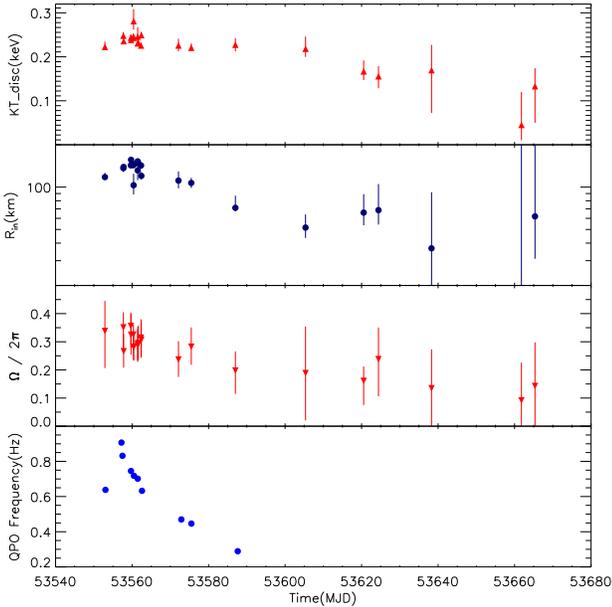}
\end{center}
\caption{The plot shows the results of the disc temperature, inner
radius and reflection evolution obtained from the {\tt diskbb+thCompml} model.  The
value of inner radius were derived from the {\tt diskbb} normalisation
assuming an inclination of $60^\circ$ and distance of 5.4kpc. As
these are poorly known, the absolute value of the radius is not
well constrained. However, trends should be robust, and clearly the
derived inner radius remains constant or mildly decreases during the
outburst. This is in sharp contrast to the expected increase from
truncated disc models. Reflection (third penal) and the QPO frequency (lower panel, see also Table \ref{bigtable}), on the other hand, decreases when leaving the peak, as expected if the disc is receding.}
\label{trend}
\end{figure}

We first concentrate on the X-ray data alone, and show the different
components present in the spectrum. We fit a power law to the
'continuum dominated' bands at 2--4~keV and 7--10~keV, and then show a
ratio of the data to this model over the entire bandpass (assuming
absorption fixed at the galactic column of $N_H=0.2\times
10^{22}$~cm$^{-2}$). Fig.~\ref{residual} shows this at different
points along the outburst (marked by arrows in Fig. 1), from the peak
to the end of the decline. Close to the peak, the continuum clearly
has a rollover at high energies, so is most likely due to thermal
Comptonisation. There is also clearly an additional soft X-ray
component, which decreases in strength as the outburst declines.
However, there are also more subtle features seen in the high
signal-to-noise PCA data, with a small excess around the iron line,
followed by a slight dip and subsequent rise to 20~keV. This is
characteristic of reflection

Thus we use a model including a disc to produce the soft X-ray
component, thermal Comptonisation of these disc seed photons to
produce the hard continuum, and reflection of the hard continuum from
the surface of the disc. We use the {\sc thcomp} model to describe the
Comptonisation, developed and tested by \citet{Zdziarski00}, extended to include its reflected continuum and
self-consistent line emission by \citet{Zycki99}.  We assume
that the reflecting material is neutral, with solar abundances and
inclined at $60^\circ$. We fix the galactic absorption column at
$N_H=0.2\times 10^{22}$~cm$^{-2}$ corresponding to an $E(B-V)=0.34$
\citep{CB07}. The results for each spectrum are detailed in Table 1.
We assess the significance of the detected disc emission by removing
the disc from the model, and fixing the seed photon temperature
instead at 0.1~eV. This results in a loss of 2 degrees of freedom, so
the disc is only significant at greater than 99 per cent confidence
for $\Delta \chi^2 > 4.61$. Thus all the spectra apart from the final
3 (26, 30 and 31) strongly require the presence of an additional soft
component. We similarly assess the significance of reflection (driven
mostly by the presence of the iron line) and find that it is required
in all the datasets apart from the final 3.

Fig. \ref{trend} shows the derived disc temperature and radius (see
Table \ref{bigtable}). The disc radius \emph{decreases} as the
outburst progresses (see also \citealt{Cabanac09}), in apparent
conflict with the truncated disc model where the inner disc should
progressively recede during the LHS decline. Yet the amount of
reflection decreases, as expected if the disc is
receding. Additionally, the low frequency QPO seen in the variability
power spectrum drops dramatically, again as expected if the truncation
radius sets the QPO frequency and is increasing (DKG07). Thus while
the behaviour of the reflected emission and QPO fit the truncated disc
models, the behaviour of the soft X-ray component does not.

\subsection{Optical data}

\begin{figure}
\begin{center}
\leavevmode \epsfxsize=6cm \epsfbox{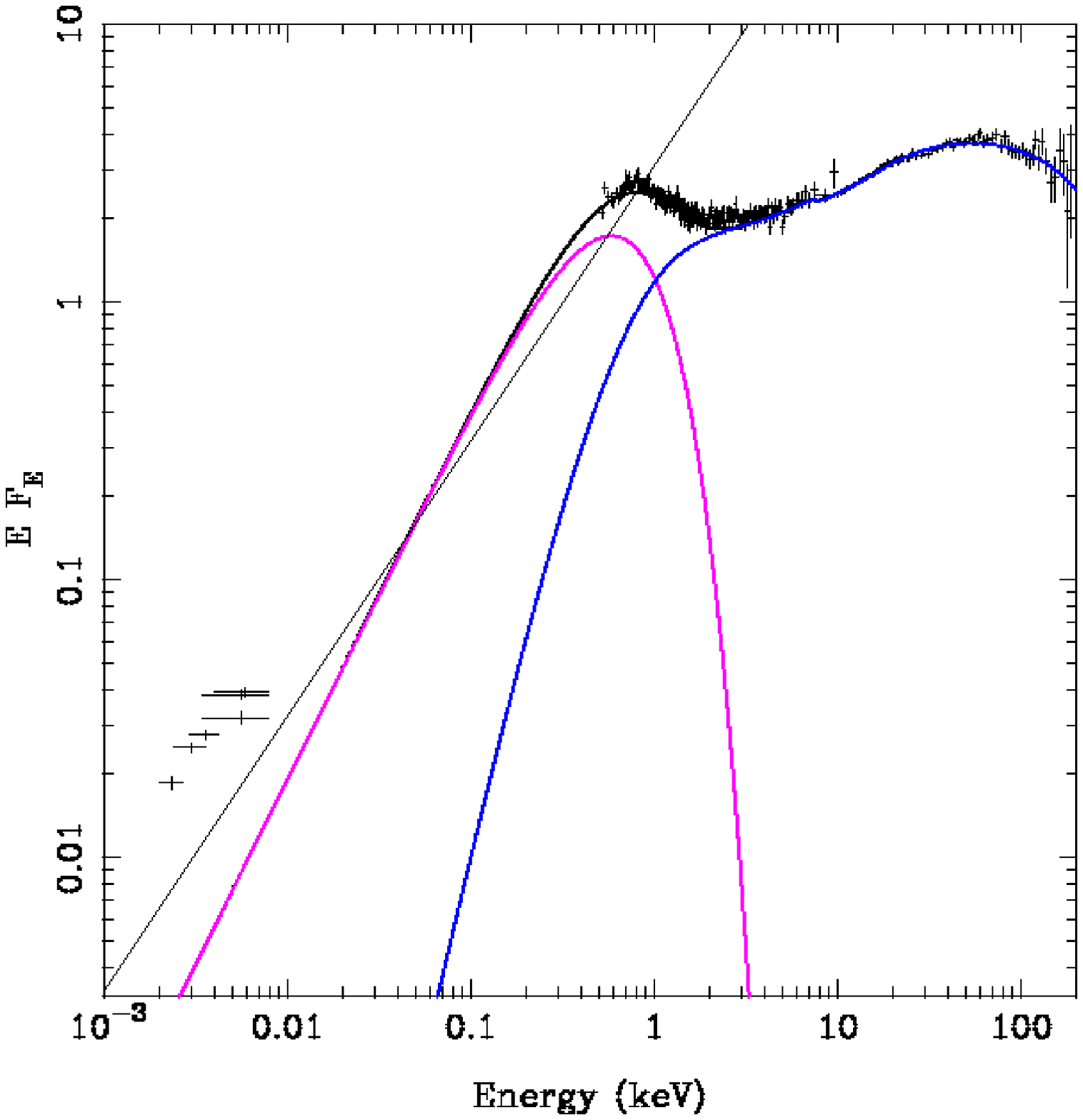}
\leavevmode \epsfxsize=6cm \epsfbox{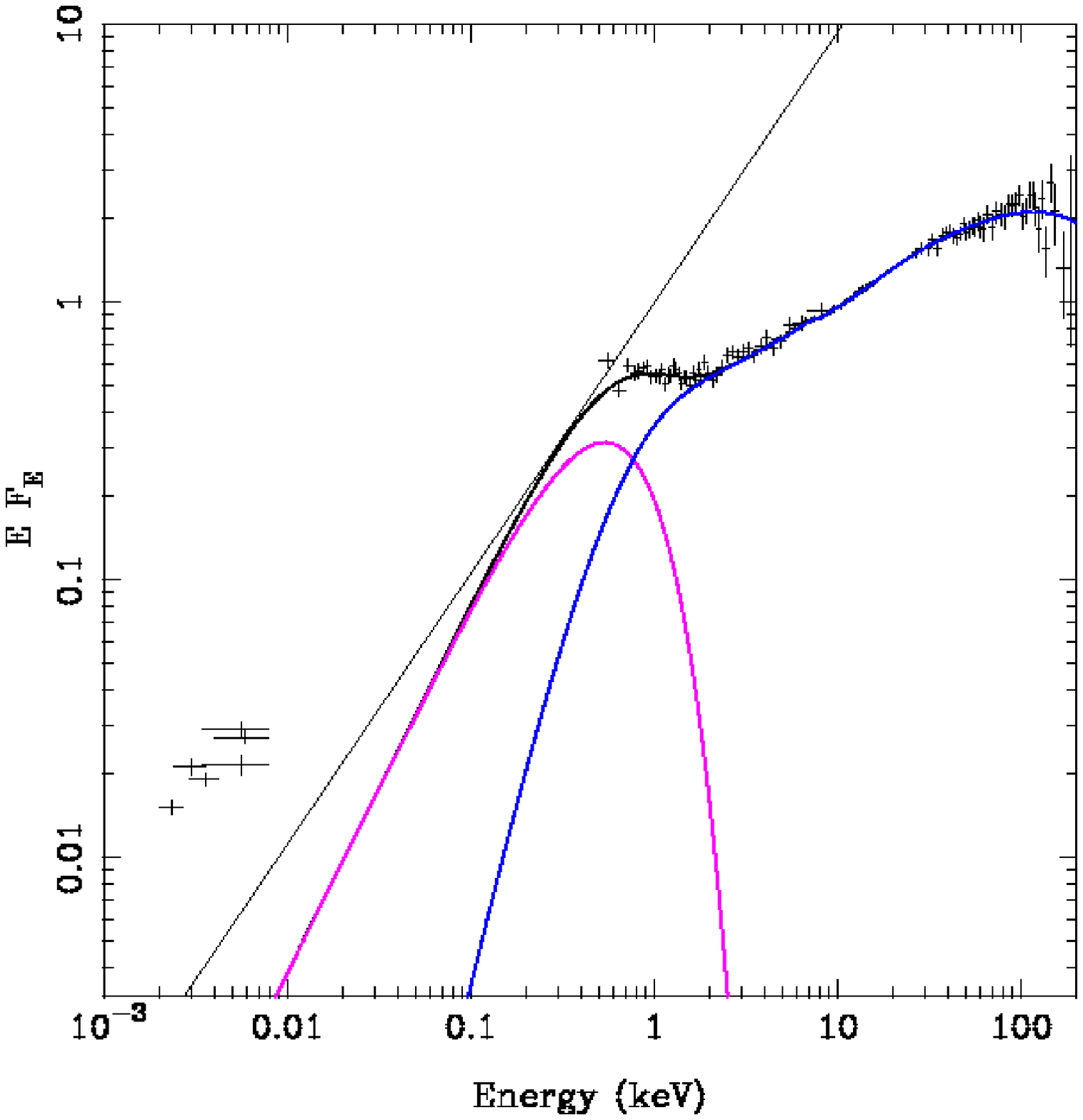}
\leavevmode \epsfxsize=6cm \epsfbox{simple30.ps}
\end{center}
\caption{The unabsorbed data and X-ray model fit ({\tt diskbb +
thCompml}) at the outburst peak (top panel, spectrum 10), midway down
the decline (middle panel, spectrum 20) and in the outburst tail
(lower panel, spectrum 30).  The magenta and blue components are disc
emission and Comptonisation in corona, respectively.}
\label{simple}
\end{figure}

We show these X-ray model fits extrapolated down to the optical/UV
data and corrected for the effects of interstellar absorption.
Fig. \ref{simple}a shows this for the outburst peak (10). This clearly
shows that extrapolating the hard X-ray power law spectrum down in
energy {\em overpredicts} the optical/UV data.  This implies that the
hard X-ray spectrum must break at UV/soft X-ray energies, supporting
our modeling of it as due to Comptonisation of seed photons from the
accretion disc.  However, extrapolating the soft X-ray disc emission
{\em underpredicts} this emission, making it clear that intrinsic
gravitational energy release in the disc is insufficient to produce
the observed optical/UV data.

In general, the optical can include additional contributions from
reprocessed emission from hard X-ray illumination of the outer disc,
as well as contributions from the jet and the companion
star(e.g. \citealt{Russel 06}).  However, SWIFT J1753.5-0127 is a
short period, low mass X-ray binary, so the companion star must
faint. Instead we try to quantify the jet contribution from observed
radio flux of $\sim 2.1$~Jy at 1.7 GHz (\citealt{Fender05}. We show
the extrapolation of this up through our optical/UV/X-ray bandpass
assuming a flat spectrum (black line on Fig. \ref{simple}a).  This is a factor
$\sim 3$ below the observed optical flux, but this is the maximum
possible jet contribution as its spectrum should break where it
becomes optically thin. It seems most likely that this break is at IR
frequencies (\citealt{Markoff01}; \citealt{Gallo07}) so the jet is also
probably negligible in the optical/UV for these data. Thus X-ray
reprocessing from hard X-ray illumination of the outer disc seems the
most likely origin for the optical/UV flux \citep{vanParadijs96}.

\begin{figure}
\begin{center}
\leavevmode \epsfxsize=8.5cm \epsfbox{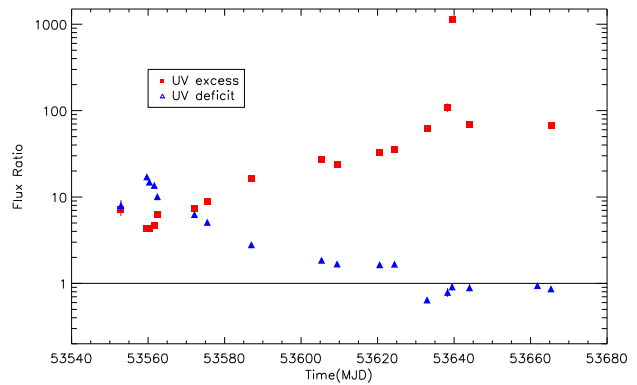}
\end{center}
\caption{The red squares show the evolution of UV excess i.e. the
observed UVW1 flux divided by the UVW1 flux predicted by the model fit
to the soft X-ray disc. This always underpredicts the observed UVW1
emission, by an amount that increases as the source fades.  The soft
X-ray disc component is not significant in the final 3 points but the
trend is already clear even without these data.  The blue triangles
show the 'UV deficit' i.e. the ratio of model to observed flux in
the UVW1 filter predicted from extrapolating the {\tt thCompml}
continuum down into the optical assuming the seed photons are at
0.1~eV rather than from the disc. Close to the peak of the outburst
this UV deficit is large, showing that the
X-rays over-predict the UV flux. Conversely, towards the end of the
outburst, the UV data lie very close to the extrapolation of the hard
X-ray flux.}
\label{uvexcess}
\end{figure}

There are additional radio observations from midway down the outburst
decline \citep{CB07}. Fig. \ref{simple}b shows this radio emission, extrapolated as
above, onto spectrum 20, the dataset closest in time to the radio
observations. Again, the optical/UV points are not fit by either the
hard X-ray power law extrapolated down (though this time the mismatch is
not so large), or by the radio emission extrapolated up, or by the
disc inferred from the soft X-ray component, as also shown in
\citet{CB07}.

Fig. \ref{simple}c shows a spectrum from the end of the outburst (30), where
there are no radio observations.  The disc emission fit to the soft
X-ray component is now very weak and cool, and its extrapolation down
to the optical lies dramatically below the observed optical/UV
emission. However, the optical/UV emission does lie remarkably close
to the extrapolated hard X-ray flux in these low-luminosity data.

We quantify this by calculating the factor by which the disc component
underpredicts the observed flux in the UVW1 filter ('UV excess':
GDP09) in each spectrum. We also calculate a 'UV deficit', which is
the factor by which the Comptonisation model {\em overpredicts} the
flux in the UVW1 filter if its seed photons put at 0.1~eV rather than
fixed to the disc temperature.  Such low energy seed photons may be
produced by cyclo-synchrotron emission by the same thermal electrons
which produce the Compton scattered emission interacting with the
tangled magnetic field in the hot flow itself (\citealt{NY95};
\citealt{MCF97}; \citealt{WZ00}). Fig. \ref{uvexcess} shows how these evolutions during
the outburst. This confirms the results seen in the three individual
spectra discussed above. The 'UV excess' (red triangles) increases
strongly with time, as the soft X-ray disc component makes less and
less contribution to the UV flux. The disc is not significantly
detected in the final three spectra, but the trend in UV excess is
already clearly apparent. Conversely, the UV deficit (blue squares)
decreases, and is roughly consistent with the observed UV flux from
spectrum 21 onwards.

\section{Irradiated Disc Model }\

We first look at the possibility that irradiation of the disc is
responsible for all the issues highlighted in the previous section.
Irradiation of the inner disc changes its derived temperature and
radius compared to that derived from fitting purely gravitational
energy release models of the disc (GDP08). Similarly, irradiation of
the outer disc can also increase the optical/UV emission from that
expected from an unilluminated disc (\citealt{vanParadijs96};GDP09)

We use a slightly modified version of the {\tt diskir} model of GDP08
so that the disc temperature and normalisation are set by the physical
variables of the mass, distance, mass accretion rate and inner and
outer radii.  This has the advantage that the outer radius is set in
terms of physical units, rather than relative to the (possibly
changing) inner radius as in {\tt diskir}. We parameterise the inner
and outer radii in terms of $R_g=GM/c^2$, and leave the inner disc
radius, $R_{d}$ as a free parameter in the fits. We fix the mass and
distance at the best (though poorly constrained) estimates of
$12~M_\odot$ and 5.4~kpc, respectively \citep{Zurita 08}.

This model uses the {\sc diskbb} parameterisation of disc temperature
with radius i.e. $T\propto r^{-3/4}$ \citep{Mitsuda84}, so does not
include a stress-free inner boundary condition or colour temperature
correction. If the disc is truncated then the lack of stress-free
inner boundary condition is probably more appropriate, but the colour
temperature correction, $f_{col}$, is still an issue.  The derived
value of the inner radius is then underestimated by a factor
$f^2_{col}\sim 2.9 $ for $f_{col}\sim 1.7$ \citep{Kubota01}. 
However, the main aim of our spectral fitting is to track {\em
changes} in the value of the inner disc radius, as its absolute value
depends on the poorly known system parameters. The key aspect of the
truncated disc model which we are testing is the prediction that the
truncation radius increases as the spectrum hardens during the
decline. Thus we are focussing on the relative values of the inner
disc radius during the decline.

The outer radius should remain constant, as should the absorption (but
see \citealt{Cabanac09}) so we constrain these by simultaneously
fitting three datasets from various points in the outburst with the
modified {\tt diskir} model, together with X-ray absorption
parameterised by {\tt wabs} and UV reddening by {\tt redden}. Assuming
standard gas to dust ratios allows us to tie $E(B-V)=1.5 \times
N_H/10^{22}$ (GDP08). This gives a best fit at $N_{H}=2.1\times
10^{21} cm^{-2}$ and $R_{out}=10^{5.25}R_{g}$, which are fixed to
these values in all fits hereafter.

\subsection{Evolution of the irradiated disc model parameters}

\begin{figure}
\begin{center}
\leavevmode \epsfxsize=8.5cm \epsfbox{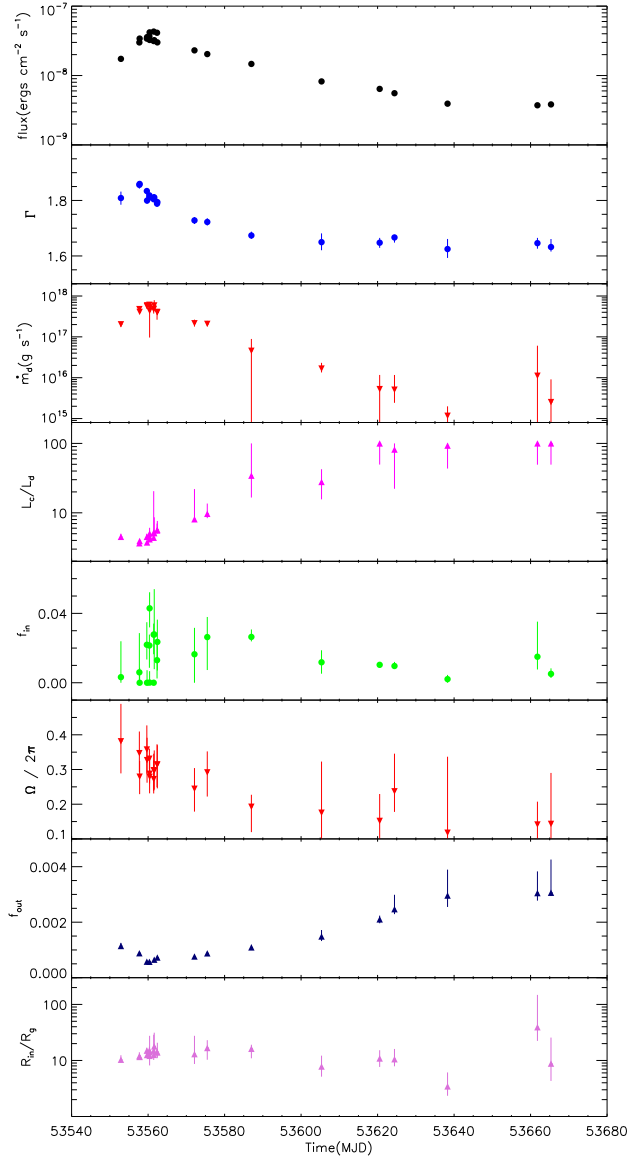}
\end{center}
\caption{Evolution of an irradiated disc model fit to the
optical/UV/X-ray spectrum. The top panel shows the bolometric flux,
taken from the unabsorbed model integrated between 0.001-100~keV.
This drops by a factor of 10 during the outburst, close to that seen
from the X-ray emission alone in Fig. 1. The second panel shows the
hard X-ray spectral index, which hardens from $1.8$ to $1.6$. The
third panel shows the mass accretion rate $\dot{m}_d$ required to
power the intrinsic (gravitational) emission from the disc luminosity,
$L_d=GM\dot{m}_d/2R_{d}$. This drops by nearly two orders of
magnitude, so the ratio of this to hard X-ray luminosity in the corona
$L_c/L_d$ increases by a factor 10 (fourth panel). The intrinsic disc
luminosity is enhanced by irradiation of the inner disc by the hard
X-ray corona. The fraction of irradiation required drops during the
decline (fifth panel) as does the amount of (neutral) reflected
emission (sixth panel). The seventh panel shows the fraction of the
bolometric flux which is required to thermalise in the outer disc in
order to make the observed optical/UV flux. This strongly increases
during the decline (see also Fig. 4). The final panel shows the
inferred inner radius of the disc. This remains remarkably constant in
this model at around $10~R_g$ (see also Fig. 2), corresponding to
$\sim 30~R_g$ after a colour temperature correction. This
lack of change in the inner disc radius is in conflict with the
truncated disc model predictions that the change in spectral hardness
(see second panel) is driven by changes in the solid angle subtended
by the disc to the hard X-ray source.  If instead there is a
separate soft X-ray component (see Fig. 8) as required in the similarly
dim low/hard state spectra from XTE J1118+480 then these data do not
constrain the disc radius.}
\label{bigplot}
\end{figure}

We fit this model to all the \emph{Swift-RXTE} datasets.  Guided by
previous low/hard state data (\citealt{Poutanen97}; GDP08), we fix the
radius of the irradiated portion of the inner disc at $1.1\times
R_{d}$. Fig. \ref{bigplot} shows the evolution of the source
parameters in this model.  The top panel gives the bolometric flux,
which varies by an order of magnitude. This is dominated by the hard
X-ray component, whose spectral photon index (second panel) decreases
from 1.8 to 1.6 during the first section of the decline, and then
stabilizes at this value hereafter.  The source did not make a
transition to the high/soft or even intermediate state, but remained
in the hard state throughout the outburst.

\begin{figure}
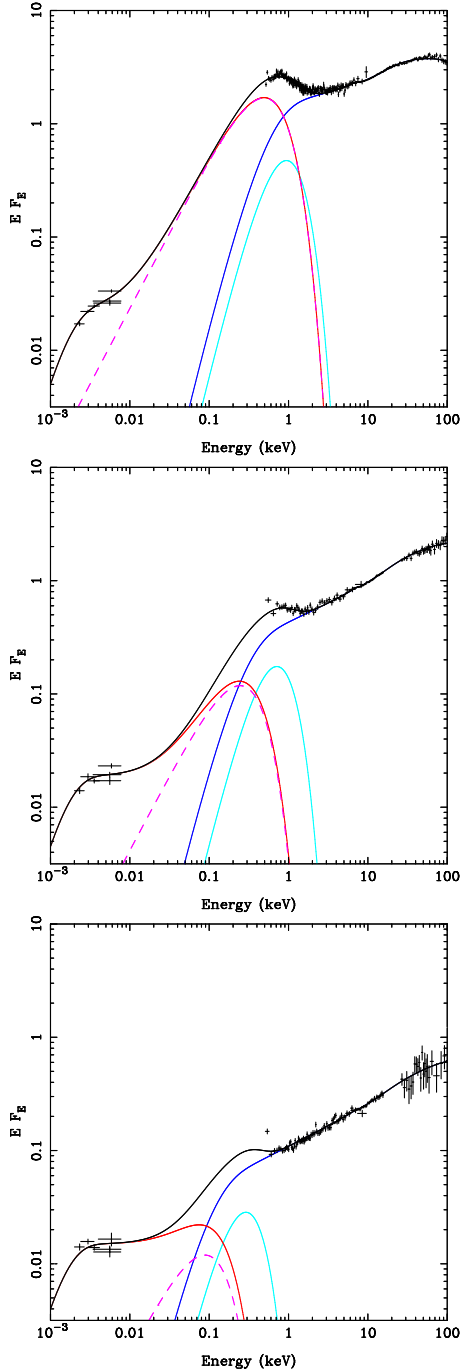

\begin{center}
\leavevmode \epsfxsize=6cm \epsfbox{comsp10.ps}
\leavevmode \epsfxsize=6cm \epsfbox{comsp20.ps}
\leavevmode \epsfxsize=6cm \epsfbox{comsp30.ps}
\end{center}
\caption{The same 3 spectra (10, 20 and 30) as in Fig. 3, fit with the
irradiated disc model and corrected (data and model) for
absorption. The magenta dashed line is the intrinsic disc emission,
the red line includes irradiation of the outer disc, and the cyan line
shows the additional flux from irradiation of the inner disc. The
Comptonised emission and its reflected spectrum are shown in
blue. While the inner disc irradiation, $f{in}$, drops, the increase
in $L_c/L_d$ means that the total disc spectrum
is more distorted  by irradiation at lower fluxes.
}
\label{irradiated}
\end{figure}

\begin{figure}
\begin{center}
\leavevmode \epsfxsize=8.5cm \epsfbox{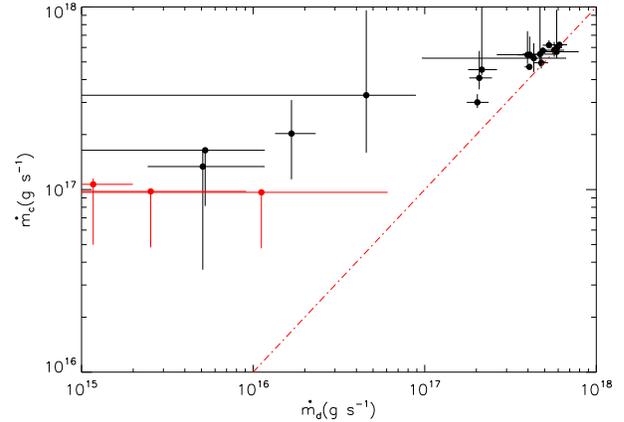}
\end{center}
\caption{Irradiated disc model mass accretion rate through the disc,
$\dot{m}_d$ versus that through the corona, $\dot{m}_c$, derived from
$(L_c/L_d) (R_{d}/R_c) \dot{m}_d$, with $R_c=3.5~R_g$ fixed by
requiring $\dot{m}_c\approx \dot{m}_d$ (shown by the red dashed line)
for the brightest spectra.  The red points show the 3 faintest spectra
where there is no significant additional soft X-ray component. Even
without these spectra the trend is clear that $\dot{m}_c$ gets
progressively larger than $\dot{m}_d$ as the source declines. This is
opposite to the expected trend from a radiatively inefficient flow.
are away from the line.}
\label{mdot}
\end{figure}

The disc spectrum forms a component which connects the soft X-ray rise
to the optical emission in this model. Part of this is powered by
gravitational energy release, and the inferred mass accretion rate
through the disc decreases by more than 2 orders of magnitude during
the decline (third panel). Thus the ratio of the hard X-ray flux to
this inferred intrinsic disc flux increases by an order of magnitude
(fourth panel).

The intrinsic gravitational energy of the disc is augmented by
thermalisation of the irradiating flux on the inner disc. The fraction
of the hard X-ray spectrum which illuminates the {\em inner} disc and
is thermalised basically decreases during the decline (fifth panel),
as expected if the inner edge of the disc is receding. Similarly, the
amount of hard X-ray reflection also declines (sixth panel).
Conversely, the fraction of hard X-rays which thermalise in the {\em
outer} disc increases by almost an order of magnitude (seventh
panel). The final panel shows the disc inner radius. This is plainly
consistent with a more or less constant value, at $\sim 10~R_g$,
(corresponding to $\sim 30~R_g$ after a colour temperature correction)
although the error bars become large at later stages of the decline.
Fig. \ref{irradiated}a-c show the corresponding fits to the data of
Figs. \ref{simple}a-c for this model.

\subsection{Inferred mass accretion rate}

We can use these derived parameters to explore how the mass accretion
rate through the disc compares to that required to power the hard
X-rays.  We calculate the mass accretion rate of the corona by:
\[
L_{c} = \frac{\eta_{corona}}{\eta_{disc}}  \frac{GM\dot{m}_{c}}{2R_{c}},
\]
where $\eta_{corona}/\eta_{disc}$ is the relative efficiency of
converting mass to radiation in a coronal flow which extends down to
$R_c$ versus that in a disc extending down to the same radius.
However, the disc itself only extends down to a radius $R_d$ so its
luminosity from gravitational energy release alone is
$L_{d} = \frac{1}{2} GM\dot{m}_{d}/R_{d}$. Thus
\[
\frac{L_c}{L_d}= \frac{\eta_{corona}}{\eta_{disc}} \frac{\dot{m}_{c}}{\dot{m}_d} \frac{R_d}{R_c}.
\]

The spectral fitting parameters give $L_c/L_d$, $\dot{m}_d$ and $R_{d}$.
Close to the peak of the outburst, even a radiatively inefficient flow
should have $\eta_{corona}/\eta_{disc}\sim 1$. Here we also expect
$\dot{m}_{c}\sim \dot{m}_{d}$, which requires $R_c$ to be around
3 times smaller than $R_d$ at this point. We can then calculate
$\dot{m}_{c}$ assuming $\eta_{corona}/\eta_{disc}$ and $R_c$ remain
constant. This is plotted against $\dot{m}_d$ in Fig. \ref{mdot}. The red line
is $\dot{m}_{c}=\dot{m}_{d}$, showing clearly that these models favour
a larger mass accretion rate through the corona than through the disc
on the decline. This conclusion is strengthened if the flow is
increasingly radiatively inefficient at lower mass accretion rates, as
the observed X-ray flux would then require an even larger coronal mass
accretion rate to power the same amount of hard X-ray emission.

This conclusion is driven by the {\em observational} requirement for a
soft X-ray component in the late state decline data.  This is weak,
but significantly present in all but the last three datasets (assuming
that the absorption column remains constant: \citealt{Cabanac09}).
It combines a low luminosity with a fairly high (soft X-ray)
temperature, which leads to the derived small radius and low mass
accretion rate.  This low mass accretion rate is then much smaller
than that required to power the observed hard X-ray emission.

We illustrate this by re-coding the irradiated disc model to force
$\dot{m}_{c}=\dot{m}_{d}$ for
$\eta_{corona}/\eta_{disc}=1$.  This forces the disc to have a higher
mass accretion rate, so it must truncate at a larger radius so as not
to overproduce the observed soft X-ray component.  Irradiation of the
inner disc should then be negligible, so we fix $f_{in}=0$ for
physical consistency, and we focus first on the X-ray data alone so we
also fix $f_{out}=0$.

\begin{figure}
\begin{center}
\leavevmode \epsfxsize=6cm \epsfbox{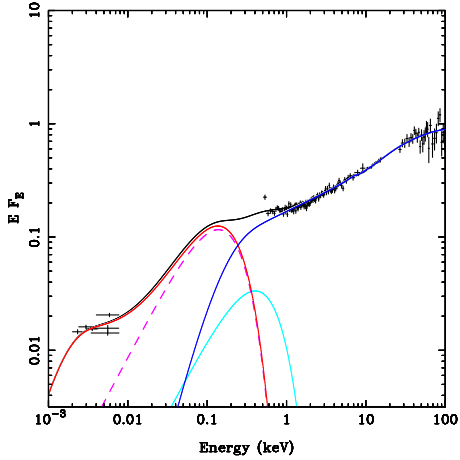}
\end{center}
\caption{Spectrum 24 fit by a model where $\dot{m}_d=\dot{m}_c$ and a
separate component (cyan, modelled with {\tt diskbb}) to fit the soft
X-ray emission. The observed strong X-ray flux requires a large mass
accretion rate through the disc, so this has to truncate at
$\sim 50~R_g$, forming the UV peak, so as not to overproduce the soft
X-ray emission.}
\label{24}
\end{figure}

We fit this to spectrum 24 (the lowest luminosity data for which the
disc is significantly detected in the soft X-ray flux), and first
focus on the X-ray data alone. The mass accretion rate through the
disc is much higher ($1.2\pm 0.1 \times 10^{17}$ g/s compared to
$1.0\times 10^{16}$ in the {\tt diskir} fits) and the disc inner
radius increases to $30~R_g$ from $\sim 5~R_g$ (i.e. $90~R_g$ and
$15~R_g$ after a colour temperature correction). This disc contributes
to the spectrum only at the softest energies of the \emph{Swift} XRT,
and its sharp rise is rather different to the more gradual curvature
seen in the soft X-ray emission. Thus this gives a significantly worse
fit ($\chi^2_\nu=626/632$ versus $600/630$, where the two extra free
parameters are $L_c/L_d$ and $f_{in}$).  We can only recover the same
quality of fit with $\dot{m}_d=\dot{m}_c$ by including an additional
soft component. This allows the disc to recede back even further, to
$\sim 50~R_g$ (i.e. $\sim 150~R_g$), so that the truncated disc makes no
contribution to the soft X-ray flux. Fig. \ref{24} shows this fit
including now the optical/UV data with $f_{out}=2\times 10^{-3}$
(similar to that derived from the original fits, see
Fig. \ref{bigplot}).

Irradiation can make such a hot, weak component if the irradiated disc
area is {\em small}. We fixed the radius of the reprocessing region at
$1.1\times R_d$ for the fits in Fig. \ref{bigplot}, but as the disc
recedes then the changing geometry means that this should also drop.
We allow this to be a free parameter with $\dot{m}_c=\dot{m}_d$ and
can recover as good a fit as before ($\chi^2=601/632$) with $\sim 1$
per cent of the bolometric flux being reprocessed in a region with
$R_{irr}=1.002 R_d$. However, illumination of such a tiny area of the
disc surface seems unreasonable from illumination by a central source.

\section{Origin of Soft X-ray Component}

\subsection{Outburst Peak}

The soft X-ray component seen at the peak of the hard state during the
outburst of SWIFT J1753.5-0127 is clearly from the disc. While the
system parameters are rather poorly known, the radii derived close to
the peak are $\sim 15~R_g$, corresponding to $\sim 45~R_g$ after a
colour temperature correction. This is consistent with the disc being
recessed back from the last stable orbit, as required for the
truncated disc/hot inner flow interpretation of the hard state. There
are two independent lines of support for this number. The first comes
from the fact that the hard X-ray spectrum is softer close to the
peak, clearly consistent with the disc providing an increasing source
of seed photons to Compton cool the hot X-ray plasma. Since most of
the gravitational potential energy to power this corona is
concentrated within $20 R_g$ then it seems most likely that the disc
extends down to similarly small radii.  This means that there can also
be irradiation of the inner disc by the hard X-ray emission, leading
to reflection and thermalisation of the incident flux. We see evidence
for both these processes in the spectrum, with the extra flux from
irradiation increasing the inferred disc radius from the value of
$\sim 11~R_g$ inferred from the simple (unirradiated) disc fits to the
X-ray spectrum, to $15~R_g$ (i,e, from $30$ to $45R_g$ after colour
temperature correction).  The second is from low frequency QPO which
is at its maximum of $\sim 0.8$~Hz in these data. This implies an
outer radius of $\sim 20-30~R_g$ assuming that the QPO is produced by
Lense-Thirring (vertical) precession of the hot flow around a
$10~M_\odot$ black hole \citep{IDF09}.

The mass accretion rate required to power the disc emission is
comparable with that required to power the coronal hard X-ray emission
assuming that the corona extends down to a radius which is $\sim 5$
times smaller than that of the disc i.e. $3.5 R_g$ for the parameters
assumed here, i.e. $\sim 9~R_g$ with a colour temperature correction.
We stress again that these system parameters are poorly known, thought
this appears quite reasonable. Much smaller radii could be potentially
feasible for a hot inner flow, irrespective of the spin of the black
hole, as numerical simulations of the MRI turbulence show that the
large scale height magnetic fields can extract energy from the
infalling material beyond the last stable orbit
(e.g. \citealt{Krolik05}).

\subsection{Late Stage Decline}

The outburst peak spectra then form a template for comparison with the
later stages of the decline, where the bolometric flux is lower by a
factor 10.  The truncated disc/hot inner flow model makes a clear
prediction that these should have a larger inner disc radius,
decreasing the importance of illumination on the inner edge of the
disc, and decreasing the amount of associated reflection.  Thus
illumination of the surface of the disc is not expected to be
important in distorting the derived inner disc radii in these
spectra. Similarly, as there is little (or no) overlap in radii between
the hot inner flow and disc, so the disc emission cannot be strongly
suppressed by Comptonisation as can be the case close to the
transition (\citealt{KD04}; \citealt{Makishima08}). Thus any disc
component seen in these spectra should give a fairly unbiased view of
the inner radius of the flow. Yet associating the observed soft X-ray
component with this disc gives derived radii which are as small, if not
smaller than, the radii seen at the outburst peak, in clear conflict
with the truncated disc models (see Figs \ref{simple} and \ref{bigplot}).

However, these radii are themselves in more subtle conflict with
the observations. The small disc has very small luminosity compared to
the coronal emission.  This requires that the coronal mass accretion
rate must be large compared to that through the disc if the corona is
powered by matter accreting through it. Thus the corona cannot be
predominantly fed by material evaporating from the inner edge of the
disc, but instead requires a completely separate coronal flow
which incorporates most of the incoming mass accretion from the
companion star. Yet it seems quite unlikely that the incoming cool
Roche lobe overflow stream would be able to form such a coronal flow at
large radii.

Instead it seems more feasible that the corona is {\em not} powered by
its own mass accretion supply, as in the truncated disc models, but is
instead powered by mass accreted through the disc, but whose energy is
released in the corona (e.g. \citealt{Svensson94}). However, this model
itself runs into difficulties since the simplest idea would be for
magnetic buoyancy to transport the energy vertically. The energy is
then released in a corona above the disc, so the corona is co-spatial
with it and illuminates it. This gives rise to a reprocessed
luminosity $L_{rep}=\frac{1}{2} (1-a) L_c$ (where $a$ is the albedo,
and the factor $\frac{1}{2}$ assumes the corona emits isotropically)
adds to the intrinsic disc luminosity \citep{HM93}.  For hard spectra,
such as those observed here, the reflection albedo $a<0.3$, hence the
disc flux should be at least $L_c/3$. Yet we observe a disc flux of
$\le L_c/20$ towards the end of the outburst. Thus the only way to
circumvent the reprocessing limits whilst having all the mass accrete
via the disc is if the energy is advected radially as well as
vertically. Then it can be released in a more centrally concentrated
region, on size scales smaller than the (small!) disc inner
radius. The alternative of having the hard X-rays be strongly beamed
away from the disc, is ruled out by the very similar dim low/hard
state spectra seen in XTE J1118+480 (\citealt{Frontera 01}; 2003;
\citealt{RMF09}), which has high inclination ($\ge 70^\circ$) so
that we must see a similar hard X-ray flux to that of the disc.

A potentially less arbitrary solution than using magnetic fields to
transport the energy in a dissipationless fashion is if the observed
soft X-ray component in the late stages of the decline is not from the
disc. The spectra clearly show that the observed soft X-ray emission
cannot produce the optical/UV emission without a large change in the
reprocessed fraction in the outer disc from that seen during the
outburst peak. Yet the outburst remained in the low/hard state, so
there is no expected change in source geometry, so the fraction of the
total flux which illuminates the outer disc should not change
dramatically. Similarly, there is no large change in spectral shape
which might produce such a large difference in thermalisation
fraction. Yet the observed soft X-ray component is a factor 10 further
below the UV emission than at the outburst peak (Fig. \ref{uvexcess}).
This could be used to argue that the real disc is truncated, as in
Fig. \ref{irradiated}, so is larger and cooler, so that the same
amount of reprocessing can produce the optical/UV.

However, it is clear that the optical/UV emission does change
character during the decline. The rapid variability of the optical
flux changes dramatically during the decline, switching to an
anti-correlation of the optical and X-ray emission \citep{Hynes
2009}. Clearly this shows that the optical is no longer made
predominantly by reprocessing of the hard X-ray flux, so it gives only
an upper limit to the irradiated disc flux.

Thus the optical emission does not trace the outer disc, so cannot
constrain our inner disc models. Instead, we use the similarly dim
low/hard state spectra from XTE J1118+480 to argue for a non-disc
origin for the soft X-ray component.  This object has an absorption
column which is an order of magnitude lower than that to SWIFT
J1753.5-0127.  This gives a correspondingly more sensitive view of the
UV and soft X-ray emission, where it is clear that there is a large,
cool disc seen in the UV and EUV bandpass (\citealt{Esin 01};
\citealt{McClintock 01}) which is completely inconsistent in
luminosity and temperature with the much weaker, higher temperature
'disc' emission seen in soft X-rays (\citealt{Frontera 01}; 2003;
\citealt{RMF09}). There are plainly {\em two} components in this
source, one which is consistent with a truncated disc, making no
impact on the soft X-ray emission, {\em and} another component which
produces weak soft X-ray flux.  This is very similar to the spectrum
of 24 as shown in Fig. \ref{24}.

Thus for the late stage decline, our modeling with an irradiated disc
is not supported by the data. The cross-correlation of the rapid
variability clearly shows that the optical/UV cannot be made by
reprocessing in the outer disc, while XTE J1118+480 clearly shows that
the soft X-rays form a separate component to the observed UV/EUV
emission from the (truncated!) disc.  What then produces the observed
soft X-ray component, and what produces the observed optical/UV emission ?

\begin{figure}
\begin{center}
\leavevmode \epsfxsize=6cm \epsfbox{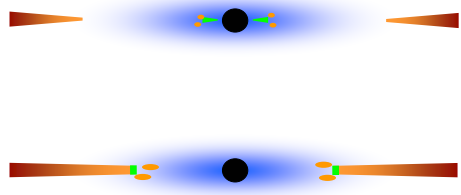}
\end{center}
\caption{Two potential geometries for a soft X-ray component
  (highlighted in green) from a
  small area, with short timescale variability dominated by
  reprocessing of the hard X-rays, and additional variability at
  longer timescales. With irradiation of a residual inner disc, the
  small area implies a small radial extent. Conversely, with
  irradiation of the inner rim of a large radius truncated disc, a
  small area implies a small vertical extent. Additional long timescale
  variability can be generated at the disc outer or inner edge,
  respectively, by clumping or turbulence.}
\label{sketch}
\end{figure}

\section{Origin of the soft X-ray component during the decline}

SWIFT J1753.5-0127 was observed by \emph{XMM-Newton} in March 2006 i.e. after
the end of the first intensive SWIFT campaign covering the outburst
decline. These data are analysed by \citet{Wilkinson09}, and have much
better signal-to-noise than those of the late decline \emph{SWIFT} spectra,
but are very similar in shape and intensity. The higher statistics in
these data mean that the soft component is {\em unambiguously}
detected, at very high significance. The same spectra model as used in
Table \ref{bigtable} ({\sc diskbb+thcomp} with absorption fixed at
$0.2\times 10^{22}$ cm$^{-2}$)gives $kT_{disc}=0.25\pm 0.01$~keV and
norm of $1080\pm 100$.

The timing analysis of \citet{Wilkinson09} showed that this soft X-ray
component variability strongly correlates with the hard X-ray
variability on short timescales i.e. is driven by
reprocessing. However, on longer timescales, there is additional
variability in the soft X-ray flux, implying that there are intrinsic
fluctuations in the disc emission as well.  The result that much of
the short term variability correlates with the hard X-rays rules out a
completely separate soft X-ray component such as the jet, while the
additional soft X-ray variability shows that the component is truly
separate as opposed to being produced from continuum curvature of the
X-ray emission or reflection (suggested by \citealt{Hiemstra 09} as a
potential origin for the soft X-rays). These variability constraints
mean that there are only two potential origins for the soft X-ray
component which arises from a small area with relatively high
temperature. Firstly, this could be emission from the surface of a
residual inner disc, forming a small ring from the last stable orbit
to some (small) outer radius. Secondly, this could be the inner face
of a truncated disc. We sketch these two possibilities in
Fig.~\ref{sketch}, and outline them below.

\subsection{Irradiation of a residual inner disc}

Evaporation of the disc by thermal conduction is a plausible mechanism
to form a truncated disc/ hot inner flow geometry
(e.g. \citealt{Liu99}).  Counterintuitively, this is not most
efficient at the smallest radii, as these also have the highest
coronal densities, so have the highest condensation rates. Evaporation
first erodes a {\em gap} in the disc, leaving a residual inner
disc. This gap expands radially, eventually giving a fully truncated
disc if the mass accretion rate is low enough\citep{Mayer07}. However,
close to the transition, the residual disc can remain. This geometry
allows the outer disc to carry all the mass accretion rate. This disc
evaporates into a corona, truncating it at some large radius into a
coronal flow which carries all the mass accretion rate. However, as
the flow accretes to smaller radii, the increasing density gives an
increasing condensation rate and some small fraction of the material
can condense out of the hot flow. This forms a cool ring at the
innermost radii, with a mass accretion rate which is only a small
fraction of the total mass accretion rate, with the rest of the
material accreting via the coronal flow\citep{Liu01}.

Reprocessing is still an issue, but if the ring is small, extending
over a very narrow range of radii, then it does not subtend a large
solid angle to the coronal X-ray emission, so could potentially
produce the large observed ratio of coronal to (inner) disc
luminosity. Some reprocessing is {\em required} in order to produce
the short term variability, but there could also be intrinsic
variability in the extent of this residual disc.

\subsection{Irradiation of the inner face of a truncated disc}

The temperature of the irradiated region is required to be
substantially hotter than that of the disc itself, requiring a very
small reprocessing area. This seems very unlikely for a central source
illuminating the top/bottom surface of thin disc.
However, the disc has some (small) half
thickness, $H_d$, so its inner rim forms a distinct, small area
$\approx 4\pi R_d H_d$ and subtends a solid angle $\approx H_d/R_d
\times 4\pi$ to the central source. The small reprocessing area found
in section 4.2 above corresponds to $H_d/R_d\sim 0.004$, predicting
$f_{in}=(1-a)H_d/R_d \approx 0.002$, where $a$ is the the reflection
albedo. This is a factor 5 smaller than $f_{in}$ derived from the
data, but potentially feasible given the large uncertainties both on
the parameters and on the modeling.

The irradiation origin then gives directly the rapid variability,
while turbulence caused by clumping instabilities on this edge could
give the required longer term additional variability.

\section{Origin of Optical/UV emission during the decline}

As noted by \citet{Motch 85} in GX339-4, the X-ray emission in the dim
low/hard state can extrapolate back quite accurately to fit the
optical spectrum (see also \citealt{CF02}; \citealt{Nowak 05}).  One
way to produce this is via a single synchrotron component from the
innermost post-shock region of the jet. This is self absorbed in the
IR, strongly suppressing the emission at lower frequencies.  As the
jet stretches out , each part of it produces a synchrotron component
which peaks at a lower frequency than those closer to the center.  All
of these make the flat spectrum seen in the radio
(e.g. \citealt{MN05}; \citealt{Maitra 09}). However, the non-thermal
synchrotron makes the optical and X-ray emission from a single
scattering in a single region, so it is hard to see how this can make
the weak anti-correlation between the optical and X-ray flux with lead
of a few hundred milliseconds which seems typical of this state
(\citealt{Motch 85}; \citealt{Kanabach01}; \citealt{Gandhi08}; \citealt{Durant08}).

Conversely, the cross-correlation signal {\em can} be explained if the
optical emission is from the jet, while the X-ray emission is from the
corona \citep{MMF04}. However, the close match of the optical
and X-ray spectra is then very unexpected if these are really from
different components.

Instead, if the spectrum is formed from thermal Comptonisation from
self-produced cyclo-synchrotron photons in the hot flow then there can
be complex time variability properties imprinted via propagating
fluctuations through an inhomogeneous flow (\citealt{KCG01};
\citealt{AU06}).  Whether these can indeed explain the
anti-correlation between optical and X-ray by such spectral pivoting
\citep{KF04} remains to be seen.

\section{Conclusions}

The combined \emph{Swift} and \emph{RXTE} observations from the black
hole transient SWIFT J1753.5-0127 give one of the best datasets to
probe the evolution of the inner edge of the accretion disc in the
low/hard state. These instruments cover the optical/UV and soft/hard
X-ray bandpasses, giving a detailed picture of the spectral evolution
during the low/hard state outburst.

We fit these with a sophisticated irradiated disc model and find that
this gives a self-consistent picture around the outburst peak. Weak
irradiation increases the inferred radius of the inner disc by a
factor $\sim 1.5$. Photons from the disc are the seeds for Compton
upscattering to produce the hard X-ray emission, and this hard X-ray
emission weakly illuminates the outer disc to produce the observed
optical/UV by reprocessing, as confirmed by the optical/X-ray
cross-correlation \citep{Hynes 2009}.

However, we find clear evidence that the model breaks down as the
source flux declines. The optical spectra require increasingly
unlikely levels of reprocessing to explain the observed emission.  A
change in origin of the optical emission is confirmed by the dramatic
change in optical/X-ray cross-correlation signal \citep{Hynes 2009}.
While the cross-correlation can be explained in models where the
optical is produced by the jet and the X-rays in a corona \citep{MMF04},
this does not explain the excellent match between the
optical and X-ray spectra. Instead it seems more likely that there is
a single component connecting the optical and X-ray spectra. The
complex cross-correlation then remains an issue especially for a
single synchrotron component from the jet, but it may potentially be
explained by thermal Comptonisation of IR cyclo-synchrotron emission in
an inhomogeneous hot flow.

More fundamentally for the focus of this paper, the soft X-ray
emission during the decline may not be thermal emission from the disc either.  If it
is, its radius does not change markedly from that seen at the outburst
peak, in clear conflict with the predictions of the truncated disc
model. But it also implies that the mass accreting through the disc is
much less than the mass accretion rate required to power the
corona. Yet it seems most likely that the mass does accrete through
the outer disc, in which case the observed weak soft X-ray disc
emission is a problem. Either the mass accretes through the disc but
most of the energy is transported vertically {\em and} radially (to
get around the reprocessing limits) by magnetic fields to power a
small, central hard X-ray corona, or the soft X-rays are from an
additional component, with the truncated disc peaking in the UV.

The lower absorption to XTE J1118+480 allows us to distinguish between
these possibilities. Here, we can see a cool component peaking in the
UV which is clearly distinct from the soft X-ray emission. The UV
component fits well to a cool disc, truncated at large radii, so the
soft X-ray component cannot be the same material. This may still be
associated with the inner disc, perhaps from irradiation of its inner
rim, or via a residual inner disc in the discontinuous disc geometry
predicted by evaporation models (e.g. \citealt{Liu01}). However, it is
also possible that the soft X-rays are instead produced in a
completely different way, such as ionised reflection from grazing
incidence angle illumination of the outer disc.
Whatever their origin, it is clear from XTE J1118+480 that the
weak soft X-ray component seen in the dim low/hard state does not
trace the inner edge of the disc, so cannot be used to constrain the
truncated disc models.

\section*{Acknowledgements}

CYC and CD would like to thank Kim Page for help in extracting the
\emph{Swift} XRT data.

%-----------------------------------------------------

\label{lastpage}

\end{document}